\documentclass[12pt]{article}

\usepackage{amsmath}
\usepackage{amssymb}
\usepackage{amsfonts}
\usepackage{latexsym}
\usepackage{color}

\catcode `\@=11 \@addtoreset{equation}{section}

\catcode `\@=12

\newcommand{\be}{\begin{equation}}
\newcommand{\en}{\end{equation}}
\newcommand{\bea}{\begin{eqnarray}}
\newcommand{\ena}{\end{eqnarray}}
\newcommand{\beano}{\begin{eqnarray*}}
\newcommand{\enano}{\end{eqnarray*}}
\newcommand{\bee}{\begin{enumerate}}
\newcommand{\ene}{\end{enumerate}}

\newcommand{\mc}{\mathcal}

\newcommand{\D}{{\mc D}}
\newcommand{\Vc}{{\mc V}}

\newcommand{\Sc}{{\cal S}}

\newcommand{\F}{{\cal F}}
\newcommand{\G}{{\cal G}}

\newcommand{\Lc}{{\cal L}}

\newcommand{\1}{1 \!\! 1}

\newcommand{\Hil}{\mc H}

\newtheorem{thm}{Theorem}

\newtheorem{prop}[thm]{Proposition}
\newtheorem{defn}[thm]{Definition}

\catcode `\@=11 \@addtoreset{equation}{section}
\catcode `\@=12

\begin{document}

\thispagestyle{empty}

\vspace*{2cm}

\begin{center}
{\Large \bf A concise review on pseudo-bosons, pseudo-fermions and their relatives}   \vspace{2cm}\\

{\large F. Bagarello}\\
  Dipartimento di Energia, Ingegneria dell'Informazione e Modelli Matematici,\\
Scuola Politecnica, Universit\`a di Palermo,\\ I-90128  Palermo, Italy\\
and I.N.F.N., Sezione di Napoli\\
e-mail: fabio.bagarello@unipa.it\\
home page: www.unipa.it/fabio.bagarello

\end{center}

\vspace*{2cm}

\begin{abstract}
\noindent We review some basic definitions and few facts recently established for $\D$-pseudo bosons  and for pseudo-fermions. We also discuss an extended version of these latter, based on biorthogonal bases, which  lives in a finite dimensional Hilbert space. Some examples are described in details.
\end{abstract}

\vspace{2cm}


\vfill


\newpage

\section{Introduction}

In this paper we review some essential definitions on the so-called $\D$-pseudo bosons ($\D$-PBs) and on pseudo-fermions (PFs). We also briefly consider a possible extension of PFs, the so-called extended PFs (EPFs), which, in a sense,  {\em interpolates} between PFs and $\D$-PBs.

$\D$-PBs arise as a modification of the canonical commutation relation (CCR) $[c,c^\dagger]=\1$, which is replaced by a similar commutation rule, $[a,b]=\1$, where $b$ is not necessarily the adjoint of $a$. Similarly, PFs are deduced replacing the canonical anticommutation relation (CAR) $\{C,C^\dagger\}=\1$ and $C^2=0$, with  $\{A,B\}=\1$ and $A^2=B^2=0$,  where again $B$ is not assumed to be the adjoint of $A$. In both these cases, biorthogonal sets of vectors can be defined, which are eigenstates of non self-adjoint number-like operators. Also, intertwining relations can be deduced and suitable metric operators can be introduced, which allow us to define different scalar products in the Hilbert space where the operators above act. Of course, this Hilbert space is infinite-dimensional in the case of $\D$-PBs, while it is just $\Bbb C^2$ for PFs. Regarding EPFs, as we will show later, they are not just the result of some generalized commutation or anticommutation relations, but are rather constructed starting from two biorthogonal bases in some finite dimensional Hilbert space $\Hil_M$, where $M=\dim(\Hil_M)$. In particular, when $M=2$, the construction produces again PFs. On the other hand, if $M>2$, we get something different. The common feature to all EPFs for any value of $M$ is the possibility of introducing and to use suitable raising and lowering operators, defined out of the bases of $\Hil_M$.

We should also mention that $\D$-PBs, PFs and EPFs appear in concrete physical models proposed along the years, see \cite{baginbagbook}
and references therein for some examples, relevant in the so-called PT-and in pseudo-hermitian quantum mechanics, \cite{ben,mosta}. We should also mention that similar deformations of the CCR, CAR etc. have been considered along the years by several other authors, see \cite{tri,yus} just to cite two such contributions.

The paper is organized as follows: in Section \ref{sectpbs} we review few facts on $\D$-PBs. PFs are considered in Section \ref{sectpfs}, while in Section \ref{sectextpfs} we consider our extended version of PFs, the EPFs, based on biorthogonal bases.   Our conclusions are given in Section \ref{sectconcl}, while some explicit physical applications are scattered all along the paper.

\section{$\D$-pseudo bosons: $\dim(\Hil)=\infty$}\label{sectpbs}

Let $\Hil$ be a given Hilbert space with scalar product $\left<.,.\right>$ and related norm $\|.\|$. Sometimes it could be useful to assume that $\Hil$ is {\em maximal}, i.e. that, given a vector $f$, in general belonging to some vector space $\Vc$ larger than $\Hil$, if $\left<f,g\right>$ is well defined for all $g\in\Hil$, then $f$ must necessarily be in $\Hil$ as well. This is, for instance, what happens in $\Lc^2(\Bbb R)$, see \cite{rudin}, while it is not true if we consider $\Hil$ to be a closed subspace of a larger Hilbert space $\Hil_{l}$, endowed with the same scalar product $\left<.,.\right>$ of $\Hil$: in this case, in fact, the fact that $\left<f,g\right>$ is well defined for all $g\in\Hil$ does not prevent $f$ to be an element of $\Hil_l$ not necessarily belonging to $\Hil$. This assumption was particularly useful in \cite{bag2016} in connection with bicoherent states.

\vspace{2mm}

Let $a$ and $b$ be two operators
on $\Hil$, with domains $D(a)$ and $D(b)$ respectively, $a^\dagger$ and $b^\dagger$ their adjoint, and let $\D$ be a dense subspace of $\Hil$
such that $a^\sharp\D\subseteq\D$ and $b^\sharp\D\subseteq\D$, where with $x^\sharp$ we indicate $x$ or $x^\dagger$. Of course, $\D\subseteq D(a^\sharp)$
and $\D\subseteq D(b^\sharp)$.

\begin{defn}\label{def21}
The operators $(a,b)$ are $\D$-pseudo bosonic  if, for all $f\in\D$, we have
\be
a\,b\,f-b\,a\,f=f.
\label{A1}\en
\end{defn}

\vspace{2mm}

{\bf Remark:--} It is not hard to imagine that the domains of the operators $a^\sharp$ and $b^\sharp$ are relevant in our context. In fact, in the particular case when $b=a^\dagger$, the pseudo-bosonic operators are really nothing but ordinary creation and annihilation operators obeying CCR, which are well known to be unbounded. Hence they can only be defined on suitable domains. The situation does not change, at least concerning this aspect, when CCR are replaced by (\ref{A1}).

\vspace{2mm}

What we want to do now is to extend the ordinary construction well known for CCR $[c,c^\dagger]=\1$ to our case. For bosons we know that a vacuum $e_0$ does exist which is annihilated by $c$, $ce_0=0$, and which belongs to the domain of all the powers of $c^\dagger$. Then we can construct a set of vectors of $\Hil$, $e_n=\frac{1}{\sqrt{n!}}(c^\dagger)^ne_0$, $n\geq0$, which are all in the domain of the number operator $N_0:=c^\dagger c$: $N_0e_n=ne_n$. The set $\F_e=\{e_n,\,n\geq0\}$ is an orthonormal (o.n.) basis for $\Hil$. If we fix $\Hil=\Lc^2(\Bbb R)$, then each $e_n$ is a well known function $e_n(x)$ in $\Sc(\Bbb R)$, the set of $C^\infty$ functions which decrease, together with their derivatives, faster than any inverse power, \cite{mess}.

When CCR are replaced by (\ref{A1}), there is no a priori reason for such a situation to remain unchanged. For this reason, we need to impose some reasonable conditions which are verified in explicit models, and which reproduce back the well known bosonic settings when $b=a^\dagger$. In particular, our starting assumptions are the following:

\vspace{2mm}

{\bf Assumption $\D$-pb 1.--}  there exists a non-zero $\varphi_{ 0}\in\D$ such that $a\,\varphi_{ 0}=0$.

\vspace{1mm}

{\bf Assumption $\D$-pb 2.--}  there exists a non-zero $\Psi_{ 0}\in\D$ such that $b^\dagger\,\Psi_{ 0}=0$.

\vspace{2mm}

It is obvious that, since $\D$ is stable under the action of the operators introduced above,  $\varphi_0\in D^\infty(b):=\cap_{k\geq0}D(b^k)$ and  $\Psi_0\in D^\infty(a^\dagger)$, so
that the vectors \be \varphi_n:=\frac{1}{\sqrt{n!}}\,b^n\varphi_0,\qquad \Psi_n:=\frac{1}{\sqrt{n!}}\,{a^\dagger}^n\Psi_0, \label{A2}\en
$n\geq0$, can be defined and they all belong to $\D$. Then, they also belong to the domains of $a^\sharp$, $b^\sharp$ and $N^\sharp$, where $N=ba$. We see that, from a practical point of view, $\D$ is the natural space to work with and, in this sense, it is even more relevant than $\Hil$. Let's put $\F_\Psi=\{\Psi_{ n}, \,n\geq0\}$ and
$\F_\varphi=\{\varphi_{ n}, \,n\geq0\}$.
It is  simple to deduce the following lowering and raising relations:
\be
\left\{
    \begin{array}{ll}
b\,\varphi_n=\sqrt{n+1}\varphi_{n+1}, \qquad\qquad\quad\,\, n\geq 0,\\
a\,\varphi_0=0,\quad a\varphi_n=\sqrt{n}\,\varphi_{n-1}, \qquad\,\, n\geq 1,\\
a^\dagger\Psi_n=\sqrt{n+1}\Psi_{n+1}, \qquad\qquad\quad\, n\geq 0,\\
b^\dagger\Psi_0=0,\quad b^\dagger\Psi_n=\sqrt{n}\,\Psi_{n-1}, \qquad n\geq 1,\\
       \end{array}
        \right.
\label{A3}\en as well as the eigenvalue equations $N\varphi_n=n\varphi_n$ and  $N^\dagger\Psi_n=n\Psi_n$, $n\geq0$. In particular, as a consequence
of these last two equations,  if we choose the normalization of $\varphi_0$ and $\Psi_0$ in such a way $\left<\varphi_0,\Psi_0\right>=1$, we deduce that
\be \left<\varphi_n,\Psi_m\right>=\delta_{n,m}, \label{A4}\en
 for all $n, m\geq0$. Hence $\F_\Psi$ and $\F_\varphi$ are biorthogonal. It is easy to see that, if $b=a^\dagger$, then $\varphi_n=\Psi_n=e_n$ (identifying $a$ with $c$), so that biorthogonality is replaced by a simpler orthonormality. Moreover, the relations in (\ref{A3}) collapse, and only one number operator exists, since in this case $N=N^\dagger$.

 The analogy with ordinary bosons suggests us to consider the following:

\vspace{2mm}

{\bf Assumption $\D$-pb 3.--}  $\F_\varphi$ is a basis for $\Hil$.

\vspace{1mm}

This is equivalent to requiring that $\F_\Psi$ is a basis for $\Hil$ as well, \cite{chri}. However, several  physical models show that $\F_\varphi$ is {\bf not} a basis for $\Hil$, but it is still complete in $\Hil$. This suggests to adopt the following weaker version of  Assumption $\D$-pb 3, \cite{baginbagbook}:

\vspace{2mm}

{\bf Assumption $\D$-pbw 3.--}  For some subspace $\G$ dense in $\Hil$, $\F_\varphi$ and $\F_\Psi$ are $\G$-quasi bases.

\vspace{2mm}
This means that, for all $f$ and $g$ in $\G$,
\be
\left<f,g\right>=\sum_{n\geq0}\left<f,\varphi_n\right>\left<\Psi_n,g\right>=\sum_{n\geq0}\left<f,\Psi_n\right>\left<\varphi_n,g\right>,
\label{A4b}
\en
which can be seen as a weak form of the resolution of the identity, restricted to $\G$. Of course, if $f\in\G$ is orthogonal to all the $\varphi_n$'s, or to all the $\Psi_n$'s, then (\ref{A4b}) implies that $f=0$. Hence $\F_\varphi$ and $\F_\Psi$ are complete in $\G$.

To refine further the structure, in \cite{baginbagbook} we have assumed that a self-adjoint, invertible, operator $\Theta$ exists, which leaves, together with $\Theta^{-1}$, $\D$ invariant: $\Theta\D\subseteq\D$, $\Theta^{-1}\D\subseteq\D$. Then we say that $(a,b^\dagger)$ are $\Theta-$conjugate if $af=\Theta^{-1}b^\dagger\,\Theta\,f$, for all $f\in\D$. This extends what happens for CCR, where $b=a^\dagger$ and $\Theta=\1$. One can prove that, if $\F_\varphi$ and $\F_\Psi$ are $\D$-quasi bases for $\Hil$, then the operators $(a,b^\dagger)$ are $\Theta-$conjugate if and only if $\Psi_n=\Theta\varphi_n$, for all $n\geq0$. Moreover, if $(a,b^\dagger)$ are $\Theta-$conjugate, then $\left<f,\Theta f\right>>0$ for all non zero $f\in \D$. The operator $\Theta$ is what sometimes is called a {\em metric operator}, and can be used to define a new scalar product on $\D$: $\left<f,g\right>_\Theta=\left<f,\Theta g\right>$, $\forall\,f,g\in\D$.

In some particular case it is possible to replace Assumption $\D$-pbw 3 above with the following stronger version:

\vspace{2mm}

{\bf Assumption $\D$-pbs 3.--}  $\F_\varphi$ is a Riesz basis for $\Hil$.

\vspace{1mm}

In this case we call our $\D$-PBs {\em regular}\footnote{Notice that the $w$ and $s$ in the Assumptions $\D$-pbs 3 and $\D$-pbw 3 stand for {\em strong} and {\em weak}, respectively. Of course, when Assumption $\D$-pbs 3 holds, Assumptions $\D$-pb 3 and $\D$-pbw3 are automatically satisfied.}.  In this case a bounded operator $S$, with bounded inverse $S^{-1}$, exists in $\Hil$, together with an orthonormal basis $\F_{\hat e}=\{\hat e_n\in\Hil,\,n\geq0\}$, such that $\varphi_n=S\hat e_n$, for all $n\geq0$\footnote{Each vector $\hat e_n$ can be identified with the $e_n$ introduced at the beginning of this section out of CCR. In fact, even if $\hat e_n\neq e_n$, we can still introduce an unitary operator $U$ such that $e_n=U\hat e_n$. Then $\varphi_n=S\hat e_n=SUe_n=S_Ue_n$, and $S_U=SU$ is still bounded with bounded inverse. Hence we assume from the very beginning that $U=\1$.}. Then, because of the uniqueness of the  basis biorthogonal to $\F_\varphi$, it is clear that $\F_\Psi$ is also a Riesz basis for $\Hil$, and that $\Psi_n=(S^{-1})^\dagger \hat e_n$. Hence, an operator $\Theta$ having the properties required above can be introduced as $\Theta:=(S^\dagger S)^{-1}$, at least if $\D$ is stable under the action of both $S^\sharp$ and $(S^\sharp)^{-1}$.  It is clear that $\Theta$ is also bounded, with bounded inverse,  self-adjoint, positive, and that $\Psi_n=\Theta \varphi_n$, for all $n\geq0$. $\Theta$ and $\Theta^{-1}$ can both be written as an (infinite) sum of rank-one operators. In fact, adopting the Dirac bra-ket notation, we have
\be
\Theta=\sum_{n=0}^\infty |\Psi_n\left>\right<\Psi_n|,\qquad \Theta^{-1}=\sum_{n=0}^\infty |\varphi_n\left>\right<\varphi_n|.
\label{add3}\en
Of course both $|\Psi_n\left>\right<\Psi_n|$ and $|\varphi_n\left>\right<\varphi_n|$ are not projection operators\footnote{Here $\left(|f\left>\right<f|\right)g=\left<f,g\right>f$, for all $f,g\in\Hil$.} since, in general, the norms of $\Psi_n$ and $\varphi_n$ are not equal to one. The series above are uniformly convergent if Assumption $\D$-pbs 3 is satisfied, while they are not, if its weaker versions, Assumption $\D$-pb 3 or $\D$-pbw 3, hold. Explicit examples of $S$ can be found in \cite{baginbagbook}. In most of these cases both $S^\sharp$ and $(S^\sharp)^{-1}$ map $\D$ into $\D$, as was assumed here. Hence, $e_n\in\D$, for all $n$. The situation is technically more complicated if Assumption $\D$-pbs 3 is replaced by its weaker version, $\D$-pbw 3.  In this case, in \cite{baginbagbook} it has been discussed that an operator $S$ and an orthonormal basis $\F_e$ can again be introduced. However, $S$ or $S^{-1}$, or both, are unbounded.

The lowering and raising conditions in (\ref{A3}) for $\varphi_n=Se_n$ can be rewritten in terms of $e_n$ as follows:
\be
S^{-1}aSe_n=\sqrt{n}\,e_{n-1},\qquad S^{-1}bSe_n=\sqrt{n+1}\,e_{n+1},
\label{a5a}\en
for all $n\geq0$. Notice that we are putting here $e_{-1}\equiv0$. Then, the first equation in (\ref{a5a}) suggests to define an operator $c$ acting on $\D$ as follows: $cf=S^{-1}aSf$. Of course, if we take $f=e_n$, we recover (\ref{a5a}). Moreover, simple computations show that $c^\dagger$ satisfies the equality $c^\dagger f=S^{-1}bS f$, $f\in\D$, which now, taking $f=e_n$, produces the second equality in (\ref{a5a}). These operators satisfy the CCR on $\D$: $[c,c^\dagger]f=f$, $\forall f\in\D$. The conclusion is that the operators $a$ and $b$ are related to a canonical pair $c$ and $c^\dagger$ by a similarity map $S$ on $\D$, which could be bounded together with its inverse, or not. We refer to \cite{baginbagbook} for a detailed mathematical treatment.

$\D$-PBs appear in several physical models arising in PT-quantum mechanics, but not only. For instance, the Hamiltonian for the non self-adjoint quantum harmonic oscillator and the Hamiltonian for the Swanson models, among others, can be written in terms of $\D$ pseudo-bosonic operators $a$ and $b$.
We refer to \cite{baginbagbook} for several applications to physics of this framework, one of which is briefly reviewed in Section \ref{sectexPBs}.  Other and more recent applications can be found in \cite{bagBS,bagPR2015,baggianf}.

\subsection{Just few drops of Bi-coherent states}\label{sectBCS}

It is well known that the bosonic annihilation operator $c$ admits a set of eigenstates labeled by a complex variable $z$. These eigenstates are
called {\em coherent states}: let $W(z)=e^{zc^\dagger-\overline{z}\,c}$, a {\em standard} coherent state is the vector
\be
\Phi(z)=W(z)e_0=e^{-|z|^2/2}\sum_{k=0}^\infty \frac{z^k}{\sqrt{k!}}\,e_k.
\label{30}\en
Here $c$ and $c^\dagger$ are operators satisfying the CCR, and $\F_e$ is the orthonormal basis related to these operators as before: $c\,e_0=0$, and $e_n=\frac{1}{\sqrt{n!}}\,(c^\dagger)^ne_0$, $n\geq0$. The vector $\Phi(z)$ is well defined, and normalized, for all $z\in\Bbb C$. This is just a consequence of the fact that $W(z)$ is unitary, or, alternatively, of the fact that $\left<e_k,e_l\right>=\delta_{k,l}$. Moreover,
\be
c\,\Phi(z)=z\,\Phi(z),\qquad\mbox{and}\qquad \frac{1}{\pi}\int_{\Bbb C}d^2z|\Phi(z)\left>\right<\Phi(z)|=\1.
\label{add1}\en
It is also well known that $\Phi(z)$ saturates the Heisenberg uncertainty relation, which will not be discussed further in this paper.

What is interesting to us here is to show that the family of vectors $\{\Phi(z),\,z\in\Bbb C\}$ can be somehow generalized in a way that preserves similar properties, and that this generalization is related to the $\D$-pb operators $a$ and $b$ considered here.

Roughly speaking, due to the relation between $(c,c^\dagger)$ with $(a,b)$ or with $(b^\dagger,a^\dagger)$, we expect we can replace $W(z)$ with one of the following operators:
\be
U(z)=e^{zb-\overline{z}\,a},\qquad V(z)=e^{za^\dagger-\overline{z}\,b^\dagger}.
\label{31a}\en
Of course, if $a=b^\dagger$, then $U(z)=V(z)$ and the operator is unitary and essentially coincide with $W(z)$, identifying $a$ with $c$. However, the case of interest here is when $a\neq b^\dagger$. This makes the situation more complicated since, in this case, neither $U(z)$ nor $V(z)$ are bounded, in general, at least when $z\neq0$. Still, in \cite{abg2015}, we have found conditions for the vectors
\be
\varphi(z)=U(z)\varphi_0,\qquad \Psi(z)=V(z)\,\Psi_0,
\label{32a}\en
to be well defined in $\Bbb C$.  The vectors $(\varphi(z),\Psi(z))$ are called bi-coherent states, for the reason that will appear clear soon. This, of course, only means that $\varphi_0$ belongs to the domain of $U(z)$, $\varphi_0\in D(U(z))$, and that $\Psi_0\in D(V(z))$, for all $z\in\Bbb C$. This was proven under some assumptions on the norms of $\varphi_n$ and $\Psi_n$:

\begin{prop}\label{prop1}
Let us assume that there exist four constants $r_\varphi, r_\psi>0$, and $0\leq \alpha_\varphi,\alpha_\psi<\frac{1}{2}$, such that $\|\varphi_n\|\leq r_\varphi^n(n!)^{\alpha_\varphi}$ and $\|\Psi_n\|\leq r_\psi^n(n!)^{\alpha_\psi}$,  for all $n\geq0$.

Then, for all $z\in \Bbb C$, $\varphi_0\in D(U(z))$ and $\Psi_0\in D(V(z))$. Moreover, $\varphi(z)\in D(a)$,  $\Psi(z)\in D(b^\dagger)$, and we have $a\,\varphi(z)=z\varphi(z)$ and $b^\dagger\Psi(z)=z\,\Psi(z)$, for all $z\in \Bbb C$. Finally, if $\F_\varphi$ and $\F_\Psi$ are biorthogonal bases for $\Hil$, then
\be
\left<f,g\right>=\frac{1}{\pi}\int_{\Bbb C}d^2z \left<f,\varphi(z)\right>\left<\Psi(z),g\right>=\frac{1}{\pi}\int_{\Bbb C}d^2z \left<f,\Psi(z)\right>\left<\varphi(z),g\right>,
\label{add2}\en
for all $f,g\in\Hil$. If $\F_\varphi$ and $\F_\Psi$ are $\D$-quasi bases, then equation (\ref{add2}) still holds, but for $f,g\in\D$.

\end{prop}

The proof of the first statement is mainly based on the uniform convergence of the series $\sum\frac{z^k}{\sqrt{k!}}\varphi_k$ and $\sum\frac{z^k}{\sqrt{k!}}\psi_k$, which is granted by the above bounds for $\|\varphi_n\|$ and $\|\Psi_n\|$.
We refer to \cite{bag2016} for more results on bi-coherent states, and to \cite{bagJMP2016} for their use in quantization.

\subsection{A physical application of $\D$-PBs}\label{sectexPBs}

The  example we want to discuss here was  proposed  in \cite{ben2,MLmiao1}, and then considered in \cite{baglat}. The
starting point is the following, manifestly non self-adjoint, hamiltonian: \be H=(p_1^2+x_1^2)+(p_2^2+x_2^2+2ix_2)+2\epsilon
x_1x_2,\label{FBML41}\en where $\epsilon$ is a real constant, with $\epsilon\in]-1,1[$, and where the self-adjoint operators $x_j$ and $p_k$ satisfy the rule
$[x_j,p_k]=i\delta_{j,k}\1$, $\1$ being the identity operator on $\Lc^2({\Bbb R}^2)$. All the other commutators are zero.

In order to rewrite $H$ in a more convenient form we introduce, see \cite{MLmiao1}:
$$ \left\{
    \begin{array}{ll}
a_1=\frac{1}{2\sqrt[4]{1+\epsilon\, \xi}}\left((ip_1+\sqrt{1+\epsilon\, \xi}\,x_1)+\xi(ip_2+\sqrt{1+\epsilon\, \xi}\,x_2)+
i\,\frac{\xi}{\sqrt{1+\epsilon\, \xi}}\right),\\
a_2=\frac{1}{2\sqrt[4]{1-\epsilon\, \xi}}\left((ip_1+\sqrt{1-\epsilon\, \xi}\,x_1)-\xi(ip_2+\sqrt{1-\epsilon\, \xi}\,x_2)-
i\,\frac{\xi}{\sqrt{1-\epsilon\, \xi}}\right),\\
b_1=\frac{1}{2\sqrt[4]{1+\epsilon\, \xi}}\left((-ip_1+\sqrt{1+\epsilon\, \xi}\,x_1)+\xi(-ip_2+\sqrt{1+\epsilon\, \xi}\,x_2)+
i\,\frac{\xi}{\sqrt{1+\epsilon\, \xi}}\right),\\
b_2=\frac{1}{2\sqrt[4]{1-\epsilon\, \xi}}\left((-ip_1+\sqrt{1-\epsilon\, \xi}\,x_1)-\xi(-ip_2+\sqrt{1-\epsilon\, \xi}\,x_2)-
i\,\frac{\xi}{\sqrt{1-\epsilon\, \xi}}\right),
       \end{array}
        \right.
$$
where $\xi$ can be $\pm 1$. We observe that $b_j\neq a_j^\dagger$, and that \be [a_j,b_k]=\delta_{j,k}\1, \label{44}\en at least formally, the other commutators being zero.
Then $H$ can be written as \be H=H_1+H_2+\frac{1}{1-\epsilon^2}\,\1,\quad
H_1=\sqrt{1+\epsilon\, \xi}(2N_1+\1), \quad H_2=\sqrt{1-\epsilon\, \xi}(2N_2+\1). \label{FBML45}\en
where $N_j:=b_ja_j$.

We can check that the two-dimensional version of the Assumptions $\D$-pb 1,
$\D$-pb 2 and $\D$-pb 3 (or $\D$-pbw 3) hold true.

For that, we first  have to find a dense subspace $\D$ of $\Lc^2({\Bbb R}^2)$ which is stable under the action of $a_j$, $b_j$ and their adjoints. Moreover
 $\D$ must also contain the two vacua of $a_j$ and $b_j^\dagger$, if they exist. Hence, from a practical point of view, it is convenient to look
 first for a solution of the equations $a_1\varphi_{0,0}(x_1,x_2)=a_2\varphi_{0,0}(x_1,x_2)=0$ and
 $b_1^\dagger\Psi_{0,0}(x_1,x_2)=b_2^\dagger\Psi_{0,0}(x_1,x_2)=0$. Using $p_j=-i\frac{\partial}{\partial x_j}$, these are simple two-dimensional
  differential equations which can be easily solved, and the results are
\be \left\{
    \begin{array}{ll}
\varphi_{0,0}(x_1,x_2)=N\exp\left\{-\frac{1}{2}\alpha_+(x_1^2+x_2^2)-k_-x_1-k_+x_2-\xi\alpha_-x_1x_2\right\},\\
\Psi_{0,0}(x_1,x_2)=N'\exp\left\{-\frac{1}{2}\alpha_+(x_1^2+x_2^2)+k_-x_1+k_+x_2-\xi\alpha_-x_1x_2\right\},
\end{array}
        \right.
        \label{FBML46}\en
where
$$
\alpha_{\pm}=\frac{1}{2}\left(\sqrt{1+\epsilon\, \xi}\pm \sqrt{1-\epsilon\, \xi}\right),\quad
k_-=\frac{-i\xi\alpha_-}{\sqrt{1-\epsilon^2}},\quad k_+=\frac{i\alpha_+}{\sqrt{1-\epsilon^2}}.
$$
$N$ and $N'$ in (\ref{FBML46}) are normalization constants, fixed requiring that $\left<\varphi_{0,0},\Psi_{0,0}\right>=1$. This scalar product is  finite since, being $\alpha_+>0$, $\varphi_{0,0}(x_1,x_2), \Psi_{0,0}(x_1,x_2)\in \Lc^2({\Bbb R}^2)$. Actually, both $\varphi_{0,0}(x_1,x_2)$ and $\Psi_{0,0}(x_1,x_2)$ belong to $\Sc({\Bbb R}^2)$, which we identify with the set $\D$ of
Section \ref{sectpbs}. This seems to be a good choice. In fact, other than having $\varphi_{0,0}(x_1,x_2),\,\Psi_{0,0}(x_1,x_2)\in\D$,
  $\D$ is also stable under the action of $a_j$, $b_j$ and of their adjoints.

  \vspace{2mm}

At this point we can construct, as usual, the functions

$$\left\{
    \begin{array}{ll}\varphi_{n_1,n_2}(x_1,x_2)=\frac{1}{\sqrt{n_1!\,n_2!}}\,b_1^{n_1}b_2^{n_2}\varphi_{0,0}(x_1,x_2),\\
\Psi_{n_1,n_2}(x_1,x_2)=\frac{1}{\sqrt{n_1!\,n_2!}}\,{a_1^\dagger}^{n_1}{a_2^\dagger}^{n_2}\Psi_{0,0}(x_1,x_2),\end{array}
        \right.$$ and the related sets
$\F_\varphi=\{\varphi_{n_1,n_2}(x_1,x_2), n_j\geq0\}$, $\F_\Psi=\{\Psi_{n_1,n_2}(x_1,x_2), n_j\geq0\}$. It is clear that both
$\varphi_{n_1,n_2}(x_1,x_2)$ and $\Psi_{n_1,n_2}(x_1,x_2)$ differ from $\varphi_{0,0}(x_1,x_2)$ and $\Psi_{0,0}(x_1,x_2)$  for some polynomial
in $x_1$ and $x_2$. Hence they are still functions in $\Sc({\Bbb R}^2)$, as expected.

\vspace{2mm} Now we have to check whether $\F_\varphi$ and $\F_\Psi$ are bases for $\Hil$ or not. This is not evident, in principle. What is
much easier to check is that these sets are both complete in $\Hil$, but we know that completeness of a certain non orthogonal set does not imply that this
set is also a basis. Following \cite{ben2} we define an unbounded, self-adjoint and invertible operator
$T=e^{\frac{1}{1-\epsilon^2}(p_2-\epsilon p_1)}$. Then, simple computations show that \be T\,H\,T^{-1}=(p_1^2+x_1^2)+(p_2^2+x_2^2)+2\epsilon
x_1x_2+\frac{1}{1-\epsilon^2}=:h.\label{FBML47}\en It is clear that, while $H\neq H^\dagger$, $h=h^\dagger$.
In \cite{baglat} it is shown that $T$ and $T^{-1}$ are densely defined. In fact, their domains $D(T)$ and $D(T^{-1})$ both contain the linear span of the eigenstates $e_{n_1,n_2}(x_1,x_2)$ of $h$, which form an o.n. basis for $\Lc^2(\Bbb R^2)$. Moreover, see again \cite{baglat}, $\F_\varphi$ and $\F_\Psi$ are $\D$-quasi bases, but they are not bases. This last claim follows from the following estimate for $\|\Psi_{n_1,n_2}\|^2=\|\varphi_{n_1,n_2}\|^2$ for large $n_1$ and $n_2$ and for non zero $\hat\delta_j$:
$$
\|\Psi_{n_1,n_2}\|^2\simeq \frac{1}{4\pi(4n_1n_2\hat\delta_1^2\hat\delta_2^2)^{1/4}}\,e^{\sqrt{8(n_1\hat\delta_1^2+n_2\hat\delta_2^2)}},
$$
which is clearly divergent for $n_j$ diverging. Here $\hat\delta_j$ are two (non zero) numbers defined along the way to construct $h$ out of $H$, \cite{baglat}. Therefore, calling $P_{n_1,n_2}$ the operator defined as
$P_{n_1,n_2}(f)=\left<\Psi_{n_1,n_2},f\right>\varphi_{n_1,n_2}$,  we conclude that $\|P_{n_1,n_2}\|=\|\varphi_{n_1,n_2}\| \|\Psi_{n_1,n_2}\|\rightarrow\infty$, for $n_j\rightarrow\infty$. Hence $\F_\varphi$ and $\F_\Psi$ cannot be bases for $\Lc^2(\Bbb R^2)$, \cite{dav}.

Let us now take $\Theta:=T^2$. It is clear that $\Theta^{-1}$ exists and that, together with $\Theta$, leaves $\D$ invariant. Moreover
$\Psi_{n_1,n_2}=\Theta\,\varphi_{n_1,n_2}$ so that, as discussed before in Section \ref{sectpbs}, $(a_j,b_j^\dagger)$ turn out to be
$\Theta$-conjugate and $\Theta$ is positive. Furthermore, the intertwining relation  $N_j f=\Theta^{-1}N_j^\dagger\Theta f$, $f\in\D$, can also be established.

\section{Pseudo fermions: $\dim(\Hil)=2$}\label{sectpfs}

In this section we consider a similar deformation of another very used and useful commutation rule, and we show which kind of results can be deduced. In particular, we will see that these results reflect what we have already found for $\D$-PBs, with the extra bonus that, due to the simplicity of the system, no assumption is required at all. The deformed CAR, by themselves, are sufficient to produce two biorthogonal sets of the two dimensional Hilbert space $\Hil=\Bbb C^2$. Of course, these two sets are automatically bases for $\Hil$: then, there is no reason for using any weak version of them. The starting point in this section is a
modification of the CAR $\{C,C^\dagger\}=C\,C^\dagger+C^\dagger\,C=\1$, $\{C,C\}=0$, between two operators, $C$ and
$C^\dagger$, acting on a two-dimensional Hilbert space $\Hil$. The CAR are replaced here by the following rules, \cite{FBpf1}: \be \{a,b\}=\1, \quad
\{a,a\}=0,\quad \{b,b\}=0, \label{FB220}\en where the interesting situation is when $b\neq a^\dagger$. Following what we have done for $\D$-PBs, the
first assumptions we might need to require are the following:

\begin{itemize}

\item {\bf p1.} a non zero vector $\varphi_0$ exists in $\Hil$ such that $a\,\varphi_0=0$.

\item {\bf p2.} a non zero vector $\Psi_0$ exists in $\Hil$ such that $b^\dagger\,\Psi_0=0$.

\end{itemize}

In fact, the existence of these two non trivial vectors is ensured by the fact that, because of (\ref{FB220}),  $\det(a)=\det(b^\dagger)=0$, necessarily. Hence the kernels of $a$ and $b^\dagger$ are non-trivial.
\vspace{3mm}

It is now possible to recover similar results as those for $\D$-PBs. In particular, we first introduce the following non zero vectors \be
\varphi_1:=b\,\varphi_0,\quad \Psi_1=a^\dagger \Psi_0, \label{FB221}\en as well as the non self-adjoint operators \be N=ba,\quad
N^\dagger=a^\dagger b^\dagger. \label{FB222}\en Of course, it makes no sense to consider $b^n\,\varphi_0$ or ${a^\dagger}^n \Psi_0$ for $n\geq2$, since all these vectors are automatically zero. This is analogous to what happens for ordinary fermions. Let now introduce the self-adjoint operators $S_\varphi$ and $S_\Psi$ via their action on a
generic $f\in\Hil$: \be S_\varphi f=\sum_{n=0}^1\left<\varphi_n,f\right>\,\varphi_n, \quad S_\Psi f=\sum_{n=0}^1\left<\Psi_n,f\right>\,\Psi_n.
\label{FB223}\en They look as the operators $\Theta$ and $\Theta^{-1}$ in (\ref{add3}). However, now there is no problem with the existence of $S_\varphi$ and $S_\Psi$, and their domains, since the sums in
(\ref{FB223}) are finite. Of course, this is extremely different from what we had to do for $\D$-PBs. Now it is very easy to
get the following results, similar in part to those for $\D$-PBs:

\begin{enumerate}

\item \be a\varphi_1=\varphi_0,\quad b^\dagger\Psi_1=\Psi_0.\label{FB224}\en

\item \be N\varphi_n=n\varphi_n,\quad N^\dagger \Psi_n=n\Psi_n,\label{FB225}\en
for $n=0,1$.

\item If the normalizations of $\varphi_0$ and $\Psi_0$ are chosen in such a way that $\left<\varphi_0,\Psi_0\right>=1$,
then \be \left<\varphi_k,\Psi_n\right>=\delta_{k,n},\label{FB226}\en for $k,n=0,1$.

\item $S_\varphi$ and $S_\Psi$ are bounded, strictly positive, self-adjoint, and invertible. They satisfy
\be \|S_\varphi\|\leq\|\varphi_0\|^2+\|\varphi_1\|^2, \quad \|S_\Psi\|\leq\|\Psi_0\|^2+\|\Psi_1\|^2,\label{FB227}\en \be S_\varphi
\Psi_n=\varphi_n,\qquad S_\Psi \varphi_n=\Psi_n,\label{FB228}\en for $n=0,1$, as well as $S_\varphi=S_\Psi^{-1}$ and the following
intertwining relations \be S_\Psi N=N^\dagger  S_\Psi,\qquad S_\varphi N^\dagger =N S_\varphi.\label{FB229}\en

\end{enumerate}

The above formulas show that (i) $N$ and $N^\dagger $ behave as (non Hermitian) fermionic number operators, having (real) eigenvalues 0 and 1; (ii) their related
eigenvectors are respectively the vectors in $\F_\varphi=\{\varphi_0,\varphi_1\}$ and $\F_\Psi=\{\Psi_0,\Psi_1\}$; (iii) $a$ and $b^\dagger$
are lowering operators for $\F_\varphi$ and $\F_\Psi$ respectively; (iv) $b$ and $a^\dagger$ are raising operators for $\F_\varphi$ and
$\F_\Psi$ respectively; (v) the two sets $\F_\varphi$ and $\F_\Psi$ are biorthonormal; (vi) the {\em very well-behaved} operators $S_\varphi$
and $S_\Psi$ map $\F_\varphi$ into $\F_\Psi$ and viceversa; (vii) $S_\varphi$ and $S_\Psi$ intertwine between operators which are not
self-adjoint, in a similar way as their counterparts do for $\D$-PBs.

The reason why we don't need any set $\D$ now, differently from what we did in Section \ref{sectpbs}, is clear. In fact, we don't need  to add any condition on the possibility of computing, for instance, $b\,\varphi_0$, as was needed for
$\D$-PBs. In fact, we can always act on a two-dimensional vector with a two-by-two matrix! Stated differently: no problem with the domains. Also, we don't need to
check (or to ask for) Assumption 3, since this is automatically satisfied: being biorthogonal, the vectors of both $\F_\varphi$ and $\F_\Psi$
are linearly independent. Hence $\varphi_0$ and $\varphi_1$ are two linearly independent vectors in a two-dimensional Hilbert space, so that $\F_\varphi$
is a basis for $\Hil$. The same argument obviously can be used for $\F_\Psi$. We will show in a moment that both these sets are also Riesz bases.
This is a consequence of the following theorem, which shows the connection between PFs and ordinary fermions:

\begin{thm}\label{FBPFtheo} Let $c$ and $T=T^\dagger$ be two operators on $\Hil$ such that $\{c,c^\dagger\}=\1$, $c^2=0$, and $T>0$. Then, defining
\be a=T\,c\,T^{-1},\quad b=T\,c^\dagger\,T^{-1},\label{FB230}\en these operators satisfy (\ref{FB220}).

Viceversa, given two operators $a$ and $b$ acting on $\Hil=\Bbb C^2$, satisfying (\ref{FB220}),  it is possible to define two operators, $c$ and $T$,
such that $\{c,c^\dagger\}=\1$, $c^2=0$, $T=T^\dagger$ is strictly positive, and (\ref{FB230}) holds.

\end{thm}

A first consequence of the proof of this theorem is that, since $\F_\varphi$ is just the image of the orthonormal basis $\F_e$ via a bounded operator, with
bounded inverse, $S_\Psi^{-1/2}$, $\F_\varphi$ is a Riesz basis, as we claimed before.  Secondly we see that, introducing the self-adjoint number operator for the fermionic operators, $N_0:=c^\dagger c$,
this can be related to both $N$ and $N^\dagger $: \be N=S_\Psi^{-1/2}N_0\,S_\Psi^{1/2}, \quad N^\dagger =S_\Psi^{1/2}N_0\,S_\Psi^{-1/2},
\label{FB231}\en which can be written as the following intertwining relations: $S_\Psi^{1/2}N_0=N^\dagger S_\Psi^{1/2}$, $S_\Psi^{1/2}N=N_0
S_\Psi^{1/2}$. Putting together these equations we can also recover (\ref{FB229}).

A similar connection can be established between bosons and $\D$-PBs. However, because of the essential role of unbounded operators in that case, not surprisingly the analogous of Theorem \ref{FBPFtheo} is much longer to state and much harder to prove. We refer to \cite{baginbagbook} for more details.

\subsection{Some examples}

In 2007, in \cite{FBtripf}, an effective non self-adjoint hamiltonian describing a two level atom interacting with an electromagnetic field was
analyzed in connection with pseudo-hermitian systems. We have shown in \cite{FBpf1} that this model can be very naturally rewritten in terms of
pseudo-fermionic operators, and that the structure previously described naturally arises. The starting point is the Schr\"odinger equation
 \be i\dot\Phi(t)=H_{eff}\Phi(t), \qquad H_{eff}=\frac{1}{2}\left(
                                                       \begin{array}{cc}
                                                         -i\delta & \overline{\omega} \\
                                                         \omega & i\delta \\
                                                       \end{array}
                                                     \right).
\label{triex}\en Here $\delta$ is a real quantity, related to the decay rates for the two levels, while the complex parameter $\omega$
characterizes the radiation-atom interaction. It is clear that $H_{eff}\neq H_{eff}^\dagger$.
It is convenient to write $\omega=|\omega|e^{i\theta}$. Then, we introduce the  operators
$$
a=\frac{1}{2\Omega}\left(
                     \begin{array}{cc}
                       -|\omega| & -e^{-i\theta}(\Omega+i\delta) \\
                       e^{i\theta}(\Omega-i\delta) & |\omega| \\
                     \end{array}
                   \right) $$ and
$$b=\frac{1}{2\Omega}\left(
                     \begin{array}{cc}
                       -|\omega| & e^{-i\theta}(\Omega-i\delta) \\
                       -e^{i\theta}(\Omega+i\delta) & |\omega| \\
                     \end{array}
                   \right).
$$
Here $\Omega=\sqrt{|\omega|^2-\delta^2}$, which we will assume here to be real and strictly positive. A direct computation shows that
$\{a,b\}=\1$, $a^2=b^2=0$. Hence $a$ and $b$ are pseudo-fermionic operators. Moreover, $H_{eff}$ can be written in terms of these operators as
$H_{eff}=\Omega\left(ba-\frac{1}{2}\1\right)$.

To recover the pseudo-fermionic structure we first need to find a non-zero vector $\varphi_0$ annihilated by $a$ and a second non-zero vector $\Psi_0$ annihilated by $b^\dagger$.
These are
$$
\varphi_0=k\left(
             \begin{array}{c}
               1 \\
               -\,\frac{e^{i\theta}(\Omega-i\delta)}{|\omega|} \\
             \end{array}
           \right),\qquad
\Psi_0=k'\left(
             \begin{array}{c}
               1 \\
               -\,\frac{e^{i\theta}(\Omega+i\delta)}{|\omega|} \\
             \end{array}
           \right),
$$
where $k$ and $k'$ are normalization constants, partially fixed by the requirement that
$\left<\varphi_0,\Psi_0\right>=\overline{k}\,k'\left(1+\frac{1}{|\omega|^2}(\Omega+i\delta)^2\right)=1$. We now also introduce the vectors
$$
\varphi_1=b\varphi_0=k\left(
             \begin{array}{c}
               \frac{i\delta-\Omega}{|\omega|} \\
               -e^{i\theta} \\
             \end{array}
           \right),\qquad
\Psi_1=a^\dagger\Psi_0=k'\left(
             \begin{array}{c}
               \frac{-i\delta-\Omega}{|\omega|} \\
               -e^{i\theta} \\
             \end{array}
           \right).
$$
It is now easy to check that $\F_\varphi=\{\varphi_0,\varphi_1\}$ and $\F_\Psi=\{\Psi_0,\Psi_1\}$ are biorthonormal bases of $\Hil$, and we can also check that
$$
H_{eff}\varphi_0=-\,\frac{\Omega}{2}\,\varphi_0,\quad H_{eff}\varphi_1=\frac{\Omega}{2}\,\varphi_1, \quad
 H_{eff}^\dagger\Psi_0=-\,\frac{\Omega}{2}\,\Psi_0,\quad H_{eff}^\dagger\Psi_1=\frac{\Omega}{2}\,\Psi_1.
$$
Therefore $H_{eff}$ and $H_{eff}^\dagger$ are isospectrals, as expected. Moreover we find
$$
S_\varphi=2|k|^2\left(
                  \begin{array}{cc}
                    1 & \frac{-i\delta}{|\omega|}\,e^{-i\theta} \\
                    \frac{i\delta}{|\omega|}\,e^{i\theta} & 1 \\
                  \end{array}
                \right),\quad
S_\Psi=\frac{|\omega|^2}{2|k|^2\Omega^2}\left(
                  \begin{array}{cc}
                    1 & \frac{i\delta}{|\omega|}\,e^{-i\theta} \\
                    \frac{-i\delta}{|\omega|}\,e^{i\theta} & 1 \\
                  \end{array}
                \right),
$$
and they turn out to be one the inverse of the other. They are also positive definite matrices, as they should.
Using now Theorem \ref{FBPFtheo}, we can use $S_\varphi^{\pm1/2}$ to define two {\em standard} fermion operators  $c$ and $c^\dagger$, and
their related number operator $N_0=c^\dagger c$, out of $a$ and $b$. Hence we easily find that
$$H_{eff}=S_\varphi^{1/2}\,h\,S_\varphi^{-1/2},$$
where $h=\Omega\left(c^\dagger c-\frac{1}{2}\1\right)$ is a self adjoint operator. This shows that the hamiltonian $H_{eff}$ is
similar to a self-adjoint operator.

More examples can be found in \cite{bagga}, where we have shown that the following Hamiltonians can be analyzed using the formalism proposed here:
\be
H_{DG}=\left(
  \begin{array}{cc}
    r\, e^{i\theta} & s\,e^{i\phi} \\
    t\,e^{-i\phi} & r\, e^{-i\theta} \\
  \end{array}
\right),
\label{m1}\en
where $r, s, t, \theta$ and $\phi$ are all real, non zero, quantities, see \cite{das}. Another interesting Hamiltonian is, \cite{gmm}
\bea
H_{GMM}=\left(
  \begin{array}{cc}
    \epsilon_1-i\Gamma_1 & \nu_0 \\
    \nu_0 &  \epsilon_2-i\Gamma_2 \\
  \end{array}
\right),
\ena
where $\Gamma_1$ and $\Gamma_2$ are positive quantities, $\epsilon_1$ and $\epsilon_2$ are reals, and $\nu_0$ is complex-valued. Another Hamiltonian which fits our framework was  considered  first in \cite{most2006}:
\bea
H_{MO}=E\left(
  \begin{array}{cc}
    \cos{\theta} & \,e^{-i\phi}\sin(\theta) \\
    \,e^{i\phi}\sin(\theta) &  -\cos{\theta} \\
  \end{array}
\right),
\label{m5}
\ena
where $\theta,\phi \in \mathbb{C} $, $\Re(\theta)\in [0,\pi)$, and $\Re(\phi)\in [0,\pi)$.  All these operators have been introduced in \cite{bagga} in connection with PT-quantum mechanics.

\section{Extended pseudo fermions: $1\leq\dim(\Hil)<\infty$}\label{sectextpfs}

In this section, rather than starting from some (anti) commutation rule, we consider an integer number greater or equal to zero,
 $M\geq0$, and a related set of $M+1$ linearly independent vectors $\F_M^{(h)}$: $\F^{(h)}_M=\{h_0^{(M)}, h_1^{(M)},\ldots,h_M^{(M)}\}$. Notice that we are not requiring this set to be o.n. However, $\F^{(h)}_M$ is surely a basis for an $M+1$-dimensional Hilbert space, $\Hil_{M}$, the linear span of these vectors.

We will show now how a rather general algebraic procedure, giving rise to some generalized raising and lowering operators, can be introduced into the game in order to produce, in  $\Hil_{M}$, a new family of vectors,  $\F^{(g)}_M=\{g_0^{(M)}, g_1^{(M)},\ldots,g_M^{(M)}\}$, which is biorthogonal to the vectors
in $\F^{(h)}_M$ and how these can be used to construct some families of intertwining operators. In this way we somehow extend what we have done in Section \ref{sectpfs}, and for this reason we call EPFs the excitations deduced here. We refer to \cite{abg2013} for more details of our construction, and for a discussion of the physical context in which EPFs have been originally introduced.

\vspace{2mm}

First of all, it is obvious that the  set $\F^{(g)}_0$ simply coincides with $\F^{(h)}_0$, except possibly for a normalization factor: indeed,
if we define $g_0^{(0)}:=\frac{1}{\|h_0^{(0)}\|^2}\,h_0^{(0)}$, then $\left\langle g_0^{(0)},h_0^{(0)}\right\rangle =1$. Both sets are bases in the 1-dimensional Hilbert space $\Hil_0$.

More interesting is the situation for $M=1$. In this case we introduce two bounded operators $a_1$ and $b_1$ via their action on the basis
$\F^{(h)}_1$: \be a_1\,h_0^{(1)}:=0,\quad a_1\,h_1^{(1)}:=h_0^{(1)}, \quad\mbox{and}\quad b_1\,h_0^{(1)}:=h_1^{(1)},\quad b_1\,h_1^{(1)}:=0.
\label{31}\en The action of $a_1$ and $b_1$ on a generic vector $f\in\Hil_1$ can easily be deduced using linearity, since $f=c_0\,h_0^{(1)}+c_1\,h_1^{(1)}$, for suitable coefficients $c_0$ and $c_1$. For instance, $a_1f=c_1\,h_0^{(1)}$. We see from (\ref{31}) that $a_1$ and $b_1$ act as lowering and raising operators on $\F^{(h)}_1$. From this definition we deduce that \be
a_1^2=0\quad b_1^2=0,\quad \{a_1,b_1\}=\1_1, \label{32}\en where $\1_1$ is the identity operator on $\Hil_1=\Bbb C^2$. These are exactly the
pseudo-fermionic anti-commutation rules considered in Section \ref{sectpfs}, see (\ref{FB220}), so that the same construction proposed there can be repeated here. The
biorthogonal basis $\F_1^{(g)}$ can be constructed by considering first a non-zero vector, $g_0^{(1)}$, orthogonal to $h_1^{(1)}$. Such a vector surely exists, since $\mathrm{dim}(\Hil_1)=2$.
Moreover, it is always possible to choose its normalization in such a way $\left\langle g_0^{(1)},h_0^{(1)}\right\rangle =1$. It is easy to
check that $g_0^{(1)}$ is the vacuum for $b_1^\dagger$: $b_1^\dagger\,g_0^{(1)}=0$. In fact, taken a generic vector $f\in\Hil_1$ and recalling
that $f$ can be written as $f=c_0\,h_0^{(1)}+c_1\,h_1^{(1)}$, for some complex $c_0$ and $c_1$, using (\ref{31}) we deduce that $\left\langle
f,b_1^\dagger \,g_0^{(1)}\right\rangle =\overline{c_0} \left\langle h_1^{(1)},g_0^{(1)}\right\rangle =0$. Then our claim follows from the
arbitrariness of $f$.

Let us now define the vector $g_1^{(1)}:=a_1^\dagger g_0^{(1)}$. Since
$\left\langle g_1^{(1)},h_0^{(1)}\right\rangle =\left\langle g_0^{(1)},a_1\,h_0^{(1)}\right\rangle =0$ and
$\left\langle g_1^{(1)},h_1^{(1)}\right\rangle =\left\langle g_0^{(1)},a_1\,h_1^{(1)}\right\rangle =\left\langle g_0^{(1)},h_0^{(1)}\right\rangle =1$, we conclude that
$\F^{(g)}_1=\{g_0^{(1)}, g_1^{(1)}\}$ is biorthonormal to $\F^{(h)}_1$. These two sets are respectively eigenstates of
$N_1^\dagger=a_1^\dagger\,b_1^\dagger$ and $N_1=b_1\,a_1$:
$$
N_1\,h_k^{(1)}=k\,h_k^{(1)},\qquad N_1^\dagger\,g_k^{(1)}=k\,e_k^{(1)},
$$
for $k=0,1$. Moreover, together they resolve the identity in $\Hil_1$:
$$
\sum_{k=0}^1\,|g_k^{(1)}\left>\right<h_k^{(1)}|=\sum_{k=0}^1\,|h_k^{(1)}\left>\right<g_k^{(1)}|=\1_1.
$$
We can also introduce two self-adjoint, positive and invertible operators
$$
S_1^{(h)}=\sum_{k=0}^1\,|h_k^{(1)}\left>\right<h_k^{(1)}|,\qquad S_1^{(g)}=\sum_{k=0}^1\,|g_k^{(1)}\left>\right<g_k^{(1)}|,
$$
or, more explicitly,
$$
S_1^{(h)}\,f=\sum_{k=0}^1\,\left\langle h_k^{(1)},f\right\rangle \,h_k^{(1)},\qquad S_1^{(g)}\,f=\sum_{k=0}^1\,\left\langle g_k^{(1)},f\right\rangle \,g_k^{(1)},
$$
for each $f\in\Hil_1$. These operators are one the inverse of the other: $S_1^{(h)}=\left(S_1^{(g)}\right)^{-1}$. Moreover, they map $\F^{(g)}_1$
into $\F^{(h)}_1$ and viceversa:
$$
S_1^{(h)}\,g_k^{(1)}=h_k^{(1)},\quad S_1^{(g)}\,h_k^{(1)}=g_k^{(1)},
$$
$k=0,1$, and they satisfy the following intertwining relations:
$$
S_1^{(g)}\,N_1=N_1^\dagger\, S_1^{(g)},\qquad N_1\, S_1^{(h)}=S_1^{(h)}\,N_1^\dagger.
$$
There is something more: since they are positive operators, the positive square roots of $S_1^{(h)}$ and $S_1^{(g)}$ surely exist. Therefore, we can define a new operator $n_1$ and new vectors $ c_k^{(1)}$ as in
$$
n_1:=\left(S_1^{(g)}\right)^{1/2}\,N_1\,\left(S_1^{(h)}\right)^{1/2},\qquad c_k^{(1)}:=\left(S_1^{(g)}\right)^{1/2}\, h_k^{(1)},
$$
$k=0,1$. It is easy to check that $n_1$ is a self-adjoint operator on $\Hil_1$, and that $\F^{(c)}_1=\{c_0^{(1)}, c_1^{(1)}\}$ is an o.n. basis
of $\Hil_1$, and that $n_1 c_k^{(1)}=k c_k^{(1)}$, $k=0,1$.

\vspace{3mm}

A similar procedure can be repeated also for $\Hil_2$. In this case, however, we lose the relations in (\ref{32}), but we still
maintain the main aspects of the functional structure. Also, this does not exclude that a different commutation or anticommutation rule can be found which returns the same results. The analysis of this particular aspect will be postponed to a future paper, \cite{bagbag}. The starting point is, as before, a basis $\F^{(h)}_2=\{h_0^{(2)}, h_1^{(2)},h_2^{(2)}\}$, not necessarily o.n. In this
case the raising and lowering operators, $b_2$ and $a_2$, are defined by an higher dimensional version of (\ref{31}):
\be
a_2\,h_0^{(2)}:=0,\quad a_2\,h_1^{(2)}:=h_0^{(2)},\quad a_2\,h_2^{(2)}:=\sqrt{2}\,h_1^{(2)},
\label{3a1}\en
and
\be
b_2\,h_0^{(2)}:=h_1^{(2)},\quad b_2\,h_1^{(2)}:=\sqrt{2}\,h_2^{(2)},\quad b_2\,h_2^{(2)}:=0.
\label{3b1}\en
In this case $a_2^3=b_2^3=0$, but $\{a_2,b_2\}\neq\1_2$. Nevertheless, if we define $N_2=b_2\,a_2$, we still get $N_2\,h_k^{(2)}=k\,h_k^{(2)}$,
$k=0,1,2$, so that the vectors $h_k^{(2)}$ are eigenstates of a number-like operator. The biorthogonal set $\F^{(g)}_2$ is now constructed
extending the previous procedure: we begin considering a vector, $g_0^{(2)}$, which is orthogonal to both  $h_1^{(2)}$ and $h_2^{(2)}$. This
vector is unique up to a normalization, which we choose  in such a way that $\left\langle g_0^{(2)},h_0^{(2)}\right\rangle =1$. We find that
$b_2^\dagger\,g_0^{(2)}=0$. Defining further $g_1^{(2)}:=a_2^\dagger\,g_0^{(2)}$ and $g_2^{(2)}:=\frac{1}{\sqrt{2}}\,a_2^\dagger\,g_1^{(2)}$,
we get
$$
\left\langle g_j^{(2)},h_k^{(2)}\right\rangle =\delta_{j,k},
$$
$j,k=0,1,2$. Hence $\F^{(g)}_2$ into $\F^{(h)}_2$ are biorthogonal bases of $\Hil_2$. The vector $g_k^{(2)}$ is eigenstate of $N_2^\dagger$:
$N_2^\dagger\,g_k^{(2)}=k\,g_k^{(2)}$, $k=0,1,2$. This can be proved by using the lowering nature of $b_2^\dagger$ on $\F^{(g)}_2$. In fact,
other than $b_2^\dagger\,g_0^{(2)}=0$, we can also check that $b_2^\dagger\,g_1^{(2)}=g_0^{(2)}$, and that
$b_2^\dagger\,g_2^{(2)}=\sqrt{2}\,g_1^{(2)}$. Two operators, $S_2^{(h)}$ and $S_2^{(g)}$, can be defined as before, and for these we can prove
exactly analogous results as those for $S_1^{(h)}$ and $S_1^{(g)}$. For instance, we can check that $S_2^{(h)}=\left(S_2^{(g)}\right)^{-1}$. Hence, what appears to be really relevant in this construction, is not
really the anticommutation rule $\{a,b\}=\1$, but the definition of the raising and lowering operators. For this reason, we call these {
particles}, {\em generalized pseudo-fermions}.

For generic $M$ we could repeat the same construction, starting from $\F^{(h)}_M$. The two operators $a_M$ and $b_M$, defined extending formulas (\ref{3a1}) and (\ref{3b1}), satisfy the following
property: $a_M^{M+1}=b_M^{M+1}=0$. As for the anti-commutator rule, we can write
$\{a_M,b_M\}=\sum_{k=0}^M\alpha_k^{(M)}\,|g_k^{(M)}\left>\right<h_k^{(M)}|$, where the coefficients can be easily found. For instance we have
$\alpha_0^{(1)}=\alpha_1^{(1)}=1$, $\alpha_0^{(2)}=1$, $\alpha_1^{(2)}=3$ and $\alpha_2^{(2)}=2$, and yet $\alpha_0^{(3)}=1$,
$\alpha_1^{(3)}=3$, $\alpha_2^{(3)}=5$ and $\alpha_3^{(3)}=3$. In matrix form we have:
$$\{a_M,b_M\}=\left(
                \begin{array}{cccccccc}
                  1 & 0 & 0 & 0 & \cdot & \cdot & 0 & 0 \\
                  0 & 3 & 0 & 0 & \cdot & \cdot & 0 & 0 \\
                  0 & 0 & 5 & 0 & \cdot & \cdot & 0 & 0 \\
                  0 & 0 & 0 & 7 & \cdot & \cdot & 0 & 0 \\
                  0 & 0 & 0 & 0 & \cdot & \cdot & 0 & 0 \\
                  0 & 0 & 0 & 0 & \cdot & \cdot & 0 & 0 \\
                  0 & 0 & 0 & 0 & \cdot & \cdot & 2M-1 & 0 \\
                  0 & 0 & 0 & 0 & \cdot & \cdot & 0 & M \\
                \end{array}
              \right),
$$
which, for instance, taking $M=1$, gives back $\{a_M,b_M\}=\1_1$: $M=1$ is the only choice which furnishes the identity in the right-hand side of $\{a_M,b_M\}$. The vectors of $\F^{(h)}_M$ are eigenstates of $N_M=b_Ma_M$, while those of the biorthogonal set
$\F^{(g)}_M=\left\{g_l^{(M)},\,l=0,1,\ldots,M\right\}$, are eigenstates of its adjoint
$N_M^\dagger$, and we have $\left\langle g_l^{(M)},h_k^{(M)}\right\rangle =\delta_{l,k}$.

We refer to \cite{abg2013} for more details, and to the role of $M$ in our construction. Here we just want to notice that EPFs have been introduced in \cite{abg2013} in the context of noncommutative quantum mechanics.

\section{Conclusions}\label{sectconcl}

We have shown how CCR and CAR can be deformed in such a way the {\em standard}, self-adjoint, number operators double and lose self-adjointness while keeping invariant their spectra. These deformations produce two biorthogonal sets of eigenvectors of these operators, which may be bases or not. Moreover, useful intertwining relations can be deduced, and other operators satisfying ordinary CCR or CAR can also be introduced, which are similar to the pseudo bosonic or to the pseudo-fermionic operators.

We also show that an interesting functional structure, based on rising and lowering operators defined on a (non) orthonormal basis, can also be constructed and that it gives rise to something which appears to be intermediate between $\D$-PBs and PFs. The price to pay is that, at a first sight, these new operators do not obey any interesting commutation relation. However, this point is currently under analysis, \cite{bagbag}, since it is known, \cite{buch}, that o.n. bases in finite dimensional Hilbert spaces (different from $\Hil_1$) can be deduced by suitable modifications of the CCR. Then it is natural to ask whether biorthogonal bases in some space $\Hil_M$ can be found by some suitable modification of the pseudo-bosonic commutation rule in (\ref{A1}).

\section*{Acknowledgements}

The author acknowledges partial support from Palermo University and from G.N.F.M. of the INdAM.

\end{document}